\documentclass[aps,prl,twocolumn,showpacs,floatfix,preprintnumbers,nofootinbib,superscriptaddress]{revtex4}

\usepackage{graphicx}
\usepackage{color}
\usepackage{natbib}
\usepackage{multirow}
\usepackage{amsmath}
\usepackage[amssymb]{SIunits}
\usepackage{epstopdf}

\newcommand{\dNeff}{\delta N_{\textrm{eff}}}
\newcommand{\Neff}{N_{\textrm{eff}}}

\newcommand{\LASAGNA}{\textsc{lasagna}}

\begin{document}

\title{Cosmology with self-interacting sterile neutrinos and dark matter - A pseudoscalar model}

\author{Maria Archidiacono}
\affiliation{Department of Physics and Astronomy,
 Aarhus University, 8000 Aarhus C, Denmark}

\author{Steen Hannestad}
\affiliation{Department of Physics and Astronomy,
 Aarhus University, 8000 Aarhus C, Denmark}
\affiliation{Aarhus Institute of Advanced Studies,
 Aarhus University, 8000 Aarhus C, Denmark}

\author{Rasmus Sloth Hansen}
\affiliation{Department of Physics and Astronomy,
 Aarhus University, 8000 Aarhus C, Denmark}
\affiliation{School of Physics, The University of New South Wales, 
Sydney NSW 2052, Australia}

\author{Thomas Tram}
\affiliation{Institut de
Th\'eorie des Ph\'enom\`enes Physiques, \'Ecole Polytechnique
F\'ed\'erale de Lausanne, CH-1015, Lausanne,
Switzerland}

\date{\today}

%\preprint{blahblah}

\begin{abstract}
Short baseline neutrino oscillation experiments have shown hints of the existence of additional sterile neutrinos in the $\electronvolt$ mass range. Such sterile neutrinos are incompatible with cosmology because they suppress structure formation unless they can be prevented from thermalising in the early Universe or removed by subsequent decay or annihilation. Here we present a novel scenario in which both sterile neutrinos and dark matter are coupled to a new, light pseudoscalar. This can prevent thermalisation of sterile neutrinos and make dark matter sufficiently self-interacting to have an impact on galactic dynamics and possibly resolve some of the known problems with the standard cold dark matter scenario. Even more importantly it leads to a strongly self-interacting plasma of sterile neutrinos and pseudoscalars at late times and provides an excellent fit to CMB data. The usual cosmological neutrino mass problem is avoided by sterile neutrino annihilation to pseudoscalars. The preferred value of $H_0$ is substantially higher than in standard $\Lambda$CDM and in much better agreement with local measurements.
\end{abstract}


\maketitle

{\em Introduction.}---
Data from a number of neutrino oscillation experiments point to the existence of a fourth, sterile neutrino with a mass around 1 $\electronvolt$ (see e.g.\ \cite{Kopp:2013vaa,Giunti:2012bc}). However, such a neutrino would be completely thermalised in the early universe through a combination of mixing and scattering \cite{Enqvist:1991qj,Hannestad:2012ky,Saviano:2013ktj}, and since there are stringent cosmological constraints on the presence of $\electronvolt$-scale neutrinos, cosmology seems at odds with the oscillation experiments unless the sterile neutrino is somehow prevented from being fully thermalised in the early Universe \cite{Archidiacono:2014apa} (see also \cite{Ade:2013zuv,Dvorkin:2014lea,Zhang:2014dxk}).

Several simple solutions exist to this problem. First of all, it is entirely possible that the underlying cosmological model differs from the standard $\Lambda$CDM universe, and in more complex models constraints on light neutrinos can be severely weakened. Even if $\Lambda$CDM does turn out to be  the correct cosmological model, oscillation data can still be made compatible provided that the sterile neutrino is at most partly thermalised or is removed by decay and/or annihilation before the rest mass becomes important for cosmological structure formation.

The generic condition for producing any given particle species is that $\Gamma > H$ at some epoch, where $\Gamma$ is the production rate and $H$ is the expansion rate of the universe. Partial thermalisation can be achieved either by lowering $\Gamma$ or by increasing $H$. Models which lower $\Gamma$ are for example models with new interactions in the sterile sector \cite{Hannestad:2013ana,Dasgupta:2013zpn,Bringmann:2013vra,Ko:2014bka}, whereas $H$ can be modified for example in models with low reheating temperature or early dark energy \cite{Rehagen:2014vna}.

In this paper we will revisit the possibility of new interactions in the sterile sector. Previous studies have all focused on interactions via a new light vector, i.e.\ a Fermi-like interaction \cite{Hannestad:2013ana,Dasgupta:2013zpn,Bringmann:2013vra,Ko:2014bka}. This has the merit of making neutrinos strongly interacting at early times while completely decoupled at late times.
Here we will investigate a new possibility - coupling neutrinos (and possibly dark matter) to a massless or very light pseudoscalar such as the majoron. Couplings to a scalar would lead to the presence of a new fifth force on which very tight bounds exist. However, since the pseudoscalar couples only to the spin of the involved particles and because macroscopic media are unpolarised no such problem exists for pseudoscalars. 

Interactions via a light pseudoscalar have the interesting property that they make the sterile neutrinos very strongly self-interacting at late times and effectively remove sterile neutrino anisotropic stress. Depending on the density of sterile neutrinos, this property could allow us to distinguish between self-interacting and free-streaming sterile neutrinos.

If dark matter couples to the same particle, it has the possibility to make the scattering cross-section strongly velocity dependent through Sommerfeld enhancement which is a desirable feature if some of the astrophysics problems related to cold dark matter are to be addressed.

{\it Model framework. ---}
Instead of constructing an explicit model we base our discussion on a simplified setup which, however, does contain all the relevant physics. The sterile neutrino is coupled to a new light pseudoscalar with mass $m_\phi \ll 1\electronvolt$ via
\begin{equation}
{\cal L} \sim g_s \phi \bar\nu \gamma_5 \nu.
\end{equation}
Later we will look at dark matter with a similar coupling to $\phi$
\begin{equation}
{\cal L} \sim g_d \phi \bar\chi \gamma_5 \chi.
\end{equation}
One important note is in order at this point: We assume the coupling to be diagonal in mass basis, such that the 3 mainly active mass states are completely uncoupled. This is the most natural assumption given that $\phi$ is associated with new physics and not related to standard model flavour.
The new interaction is also felt partly by the active Standard Model neutrinos, although suppressed by the mixing angle. Limits from cosmology~\cite{Archidiacono:2013dua} are not relevant, as the active neutrino mass states do not feel the new coupling, but constraints from supernovae~\cite{Farzan:2002wx, Kachelriess:2000qc} and laboratory measurements~\cite{Bernatowicz:1992ma} do apply. The supernova bounds are derived by requiring that the pseudoscalars do not carry away a significant amount of the energy released by the supernova which results in a bound on the coupling of electron neutrinos to the pseudoscalar~\cite{Farzan:2002wx}, $g_e \lesssim 4 \times 10^{-7}$. If the coupling becomes much larger, the pseudoscalars will be caught in the supernova, and the bound disappears again. However, almost all of these values are excluded by laboratory experiments~\cite{Bernatowicz:1992ma}, and we will only consider the supernova limit here. For the sterile neutrinos, the bound on $g_e$ comes from the process $\nu_e \nu_e \rightarrow \phi$, and it translates into the bound $g_s \lesssim g_e / \sin^2\theta_s = 3 \times 10^{-5}$, using $\sin^22\theta_s \sim 0.05$ from the short baseline experiments~\cite{Kopp:2013vaa,Giunti:2012bc}, where $\theta_s$ is a mixing angle representative for $(\nu_e, \nu_s)$ mixing or $(\nu_\mu, \nu_s)$ mixing. Although supernovae give the strongest bounds on the coupling strength, they are quite dependent on details in the assumptions about the supernova, and it might be more appropriate to quote the bound as $g_s \lesssim 10^{-4}$.

Let us now go through the implications of this new interaction, first for the sterile neutrinos and subsequently for the dark matter.

{\it Sterile neutrinos. ---}
The new interaction introduces a matter potential for sterile neutrinos of the form~\cite{Babu:1991at,Enqvist:1992}
\begin{equation}\label{eq:Vs}
V_s(p_s) = \frac{g_s^2}{8 \pi^2 p_s} \int p dp \left(f_\phi + f_s \right),
\end{equation}
where $f_\phi$ is the Bose-Einstein distribution for the pseudoscalar and $f_s$ is the distribution for the sterile neutrinos (see e.g.\ 
\cite{Enqvist:1990ad,McKellar:1992ja,Sigl:1992fn,Enqvist:1991qj,Stodolsky:1986dx} for a discussion of matter potentials in the standard model).
Note that the potential in Eq.~(\ref{eq:Vs}) arises from bubble diagrams and is non-zero even in a CP-symmetric medium.

Before proceeding with a quantitative calculation we can estimate how large $g_s$ needs to be in order to block thermalisation.
Consider a scenario with thermal $\phi$ and $\nu_s$ distributions characterised by a common temperature $T$. The potential is then
\begin{equation}
V_s \sim 10^{-1} \, g_s^2 T.
\end{equation}
In the absence of non-standard effects, the sterile neutrinos would be thermalised through oscillations at $T \sim 10 (\delta m^2/\electronvolt^2)^{1/6} \mega\electronvolt \sim 10 ~\mega\electronvolt$~\cite{Enqvist:1991qj}. To prevent this effect, we need to suppress the mixing angle in matter, $\theta_m$ as the production rate is proportional to $\sin^2 2\theta_m$. This is achieved if the matter potential dominates the energy difference associated with vacuum oscillation, i.e.
\begin{equation}
V \gtrsim \frac{\delta m_{\nu_s}^2}{2E} \sim \frac{\delta m_{\nu_s}^2}{T},
\end{equation}
prior to neutrino decoupling at $T \sim 1~\mega\electronvolt$  so that
\begin{equation}
g_s^2 \gtrsim 10 \frac{\delta m_{\nu_s}^2}{T^2} \sim  10^{-11}.
\label{eq:estimate}
\end{equation}
So a priori we expect that a value of $g_s \sim 3 \cdot 10^{-6}$ is sufficient to block thermalisation. It should be noted here that since the pseudoscalar coupling is diagonal in mass basis the active state feels an additional matter potential associated with the $\phi$ background. The magnitude of the potential felt by the active state is approximately $V \sim \sin^2(\theta_s) V_s \sim 0.01 V_s$. The only effect is a minute shift in the effective mass difference, corresponding to a shift of less than one percent in $g_s$.

{\it Thermal history of the sterile neutrino. ---}
The sterile neutrino can in principle be thermalised via incoherent processes such as $\phi \phi \leftrightarrow \bar \nu_s \nu_s$, assuming that there is a pre-existing background of $\phi$. 
The thermally averaged cross section in the highly relativistic limit can be calculated to be~\cite{Dolgov:1996fp}
\begin{equation}
  \label{eq:pair}
\langle \sigma |v| \rangle = \frac{g_s^4}{8 \pi T^2}  .
\end{equation}
Conservatively assuming that $g_s \sim 10^{-4}$ we find that $\nu_s$ and $\phi$ come into equilibrium at a temperature of $T \sim 1~\giga\electronvolt$, i.e.\ significantly before the oscillation process becomes important~\cite{Mirizzi:2012we}. However, since the dark sector is decoupled it does not share the entropy transfer to the standard model particles, and the end result is that when oscillations become important at $T \sim 10 ~\mega\electronvolt$, a low-temperature background of $\phi$ and $\nu_s$ exists.
However if $g_s$ is significantly lower no thermalisation occurs before the oscillation period.

{\it Results and numerical implementation. ---}

We compute the thermalisation process by solving the Quantum Kinetic Equations (QKEs) for a simplified two-neutrino framework with oscillations between $\nu_\mu$ and $\nu_s$ using a modified version of our public code \LASAGNA{}~\cite{Hannestad:2013pha}. The formulation of the QKEs~\cite{Kainulainen:2001cb, Hannestad:2012ky, Barbieri:1990vx, Enqvist:1990ad, McKellar:1992ja, Sigl:1992fn, Enqvist:1991qj} is based on an expansion of the density matrices, $\rho$, in terms of $P_a$, $P_s$, $P_x$, and $P_y$
\begin{equation*}
  \rho = \frac{1}{2}f_0
  \begin{pmatrix}
    P_a & P_x-iP_y\\
    P_x+iP_y & P_s
  \end{pmatrix},
\end{equation*}
where $f_0$ is the Fermi-Dirac distribution function. 
The QKEs are now
\begin{align*}
  \dot P_a &= V_x P_y + \Gamma_a \left[ 2 - P_a\right],\\
  \dot P_s &= -V_x P_y + \Gamma_s \left[2 \frac{f_{\textrm{eq},s}(T_{\nu_s},\mu_{\nu_s})}{f_0} - P_s\right],\\
  \dot P_x &= -V_z P_y - D P_x,\\
  \dot P_y &= V_z P_x - \frac{1}{2} V_x (P_a-P_s) - D P_y .
\end{align*}
Here, the potentials are given by
\begin{align*}
  V_x &= \frac{\delta m_{\nu_s}^2}{2p} \sin 2\theta_s,\\
  V_z &= -\frac{\delta m_{\nu_s}^2}{2p} \cos 2\theta_s - \frac{14\pi^2}{45\sqrt{2}} p \frac{G_F}{M_Z^2} T^4 n_{\nu_s} + V_s ,
\end{align*}
where $p$ is the momentum, $G_F$ is the Fermi coupling constant, $M_Z$ is the mass of the Z boson, and $n_{\nu_s} = \int f_s d^3p/(2\pi)^3$ is the number density of sterile neutrinos. For the repopulation of the active neutrinos, we use the expression
\begin{equation*}
  \Gamma_a = C_\mu G_F^2 p T^4, \quad C_\mu \approx 0.92 .
\end{equation*}
For the sterile neutrino redistribution, we choose $T_{\nu_s}$ and $\mu_{\nu_s}$ to conserve energy and number density, when $f_{\textrm{eq},s} = (e^{p/T_{\nu_s} - \mu_{\nu_s}/T_{\nu_s}}+1)^{-1}$, and we approximate the rate by
\begin{equation}
\Gamma_s = \frac{g_s^4}{ 4 \pi T_{\nu_s}^2} n_{\nu_s}.
\end{equation}
Finally, we approximate the damping term by $D = \frac{1}{2}(\Gamma_a + \Gamma_s)$.

\begin{figure}%
\includegraphics[width=\columnwidth]{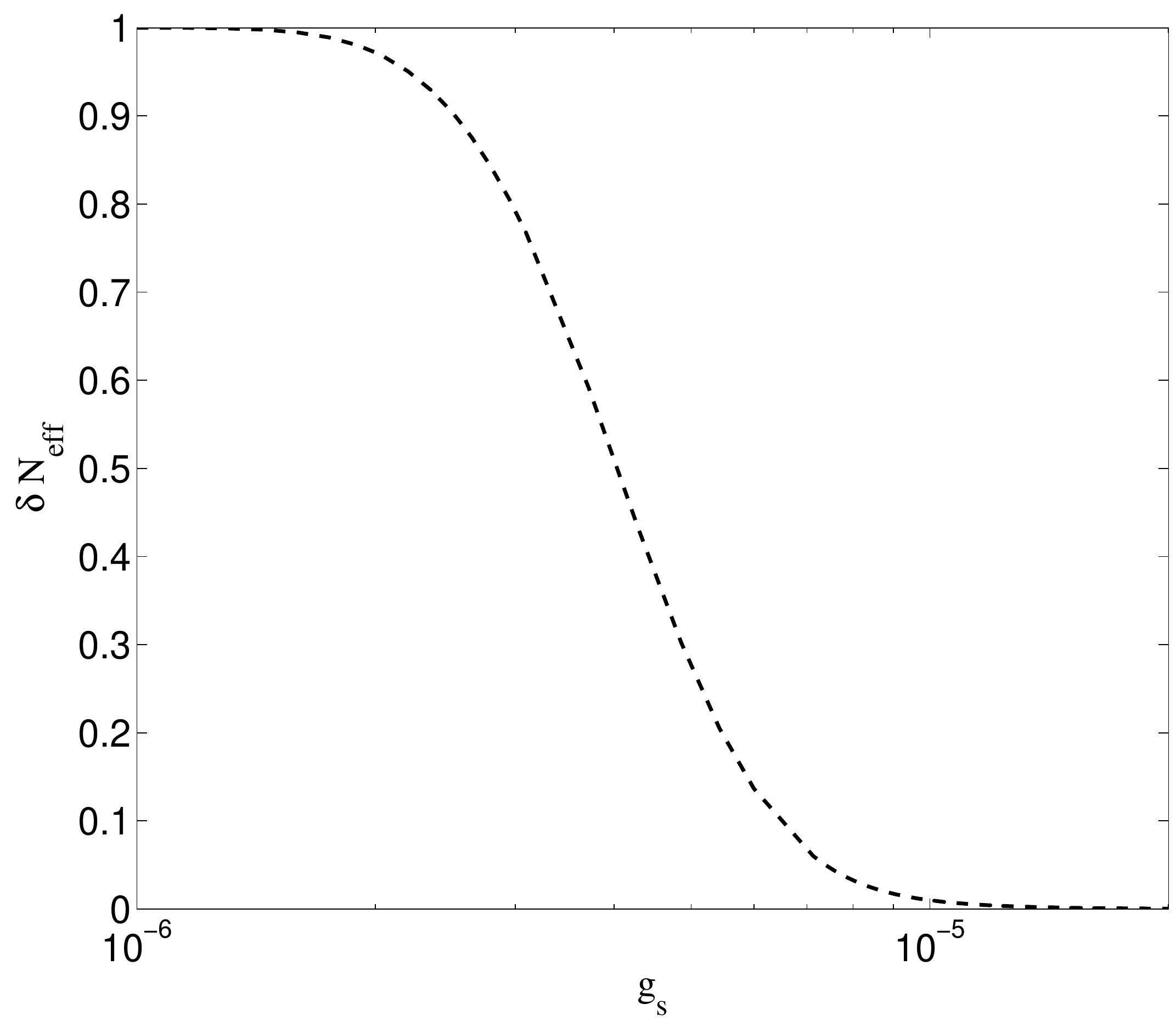}%
\caption{The contribution of the sterile neutrino to the relativistic energy density $\dNeff = \Neff - 3$ as a function of the coupling parameter $g_s$.}%
\label{fig:Neff}%
\end{figure}

We compute the sterile neutrino contribution to the potential in Eq.~\eqref{eq:Vs} from the actual numerical distribution. The contribution from the $\phi$-background is computed analytically assuming that the $\phi$-particles were produced thermally above a $\tera\electronvolt$. They will then follow a Bose-Einstein distribution with a reduced temperature of 
\begin{equation}
T_\phi = \left( \frac{g_\star(T_\gamma)}{g_\star(1 \tera\electronvolt)} \right)^{\frac{1}{3}} T_\gamma \simeq \left( \frac{10.75}{106.7} \right)^{\frac{1}{3}} T_\gamma \simeq 0.47 T_\gamma,
\end{equation}
where the approximation is valid in the temperature range of interest. We are ignoring momentum transfer between the sterile neutrinos and the pseudoscalars for simplicity, but we suspect that including it would have a negligible effect on our results. When sterile neutrinos are produced, they will create non-thermal distortions in the sterile neutrino distribution, and the sterile neutrino spectrum might end up being somewhat non-thermal. In Fig.~\ref{fig:Neff} we show the final contribution to the energy density $\Neff$
\begin{equation*}
 \Neff \equiv \frac{\rho_{\nu_a}+\rho_{\nu_s}}{\rho_{\nu_0}}\text{, where }\rho_{\nu_0}\equiv\frac{7}{8}\left(\frac{4}{11}\right)^{4/3}\rho_\gamma
 \end{equation*}
from a sterile neutrino with mixing parameter $\sin ^2 2\theta_s = 0.05$ and $m_{\nu_s} = 1~\electronvolt$, close to the best fit value from neutrino oscillation data \cite{Kopp:2013vaa,Giunti:2012bc}. The transition from full thermalisation to zero thermalisation happens in the region $10^{-6} < g_s < 10^{-5}$, confirming the simple estimate in Eq.~(\ref{eq:estimate})
\footnote{Note that in the absence of a pre-existing population of $\phi$ and $\nu_s$, sterile neutrino production would still be suppressed for the same values of $g_s$ as soon as a small amount of $\nu_s$ has been produced through oscillations. The assumption is thus not crucial to the scenario.}.

{\it Late time phenomenology. ---}
In a recent paper by Mirizzi et al.~\cite{Mirizzi:2014ama} it was pointed out that even if strong self-interactions prevent thermalisation of the sterile neutrino before active neutrino decoupling it will eventually be almost equilibrated by oscillations at late times. This leads to a scenario in which active and sterile neutrino distributions have similar temperatures and both contribute to the combined $\Neff$.
Even if early thermalisation is prevented this still leads to a sterile neutrino population with a temperature only slightly below that of standard model active neutrinos and therefore the usual cosmological neutrino mass bound still applies to this model.

However, unlike the previously studied Fermi-like interaction, sterile neutrinos and pseudoscalars interact via a variety of $2 \leftrightarrow 2$ processes which in general have a scattering rate of order $\Gamma \sim g_s^4 T$ because there is no mass scale involved. This is true for example for the pair annihilation process $\nu_s \bar\nu_s \to \phi \phi$ where we already found the thermally averaged cross section to be  $\langle \sigma |v| \rangle = g_s^4/(8 \pi T^2)$ in the relativistic limit, implying a reaction rate $\Gamma = \langle \sigma |v| \rangle n_{\nu_s} \approx 3.6 \times 10^{-3} g_s^4 T$. % = 3 \zeta(3) g_s^4 T /(32 \pi^3)
This should be compared to the Hubble expansion rate $H \sim 10 T^2/m_{\rm Pl}$. 
As long as $g_s \gtrsim 10^{-6}$ the $\nu_s - \phi$ plasma becomes strongly self-interacting before the sterile neutrinos become non-relativistic around recombination. The strong self-interactions of the combined fluid leads to a complete absence of free-streaming and in turn an absence of anisotropic stress in this component.

The scenario where all neutrinos are strongly interacting is strongly disfavoured by current data (see e.g.\ \cite{Raffelt:1987ah,AtrioBarandela:1996ur,Beacom:2004yd,Hannestad:2004qu,Cirelli:2006kt,Friedland:2007vv,Basboll:2008fx,Cyr-Racine:2013jua,Archidiacono:2013dua} for discussions of self-interacting neutrinos and cosmic structure formation). However, this is not necessarily true for models in which standard model neutrinos are free-streaming, and the interaction is confined to the sterile sector. We note here that since the pseudoscalar coupling is diagonal in mass basis it does not induce self-interactions in the three active mass states.

We also note that the rest mass constraint does not apply to this model, if we require $g_s \gtrsim 10^{-6}$: As soon as sterile neutrinos become non-relativistic they annihilate into $\phi$. This annihilation has two immediate effects. It leads to an overall increase in the energy density of the $\nu_s-\phi$ fluid, and it leads to a temporary decrease in the equation of state parameter for the fluid. Both of these effects were discussed in detail in \cite{Hannestad:2004qu}.

We have performed a study of how this model is constrained by current CMB data through an MCMC sampling of the cosmological parameter space performed with \texttt{CosmoMC} \cite{Lewis:2002ah} and using CMB data from the Planck mission as well as CMB polarisation data from the WMAP satellite \cite{Ade:2013kta} (we refer to this data combination as ``Planck+WP''). We describe the neutrino sector by the overall energy density after thermalisation, $\Neff$ and assume a sterile mass of $1~\electronvolt$. We assume complete equilibration between all species between the thermalisation scale at a few $\mega\electronvolt$ and the CMB scale ($T \sim 1~\electronvolt$), so that the energy density in the active sector is $21/32 \Neff$ with the remaining $11/32 \Neff$ is in the $\nu_s\phi$ fluid.

\begin{figure}[t]%
\includegraphics[width=\columnwidth]{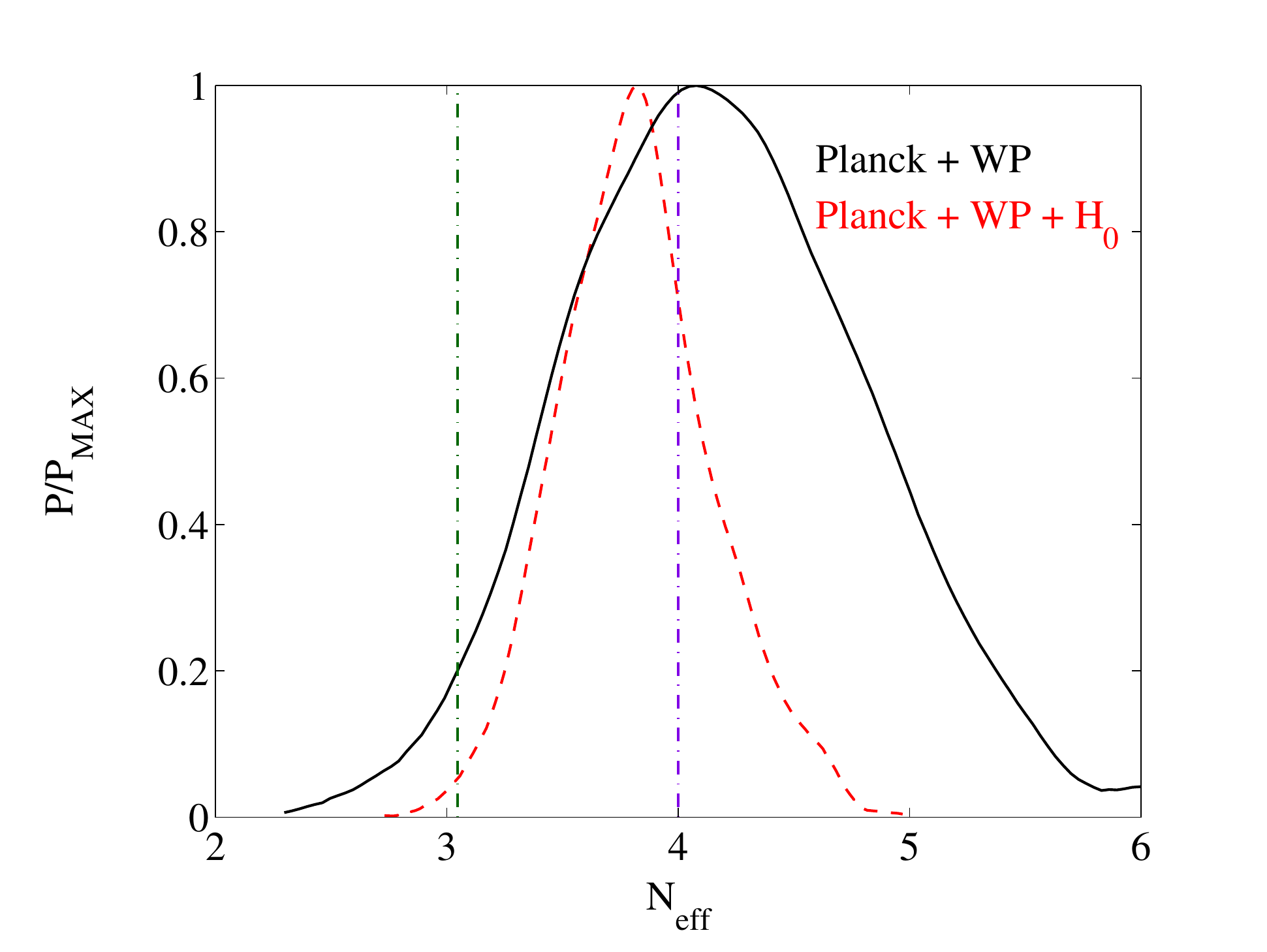} \includegraphics[width=\columnwidth]{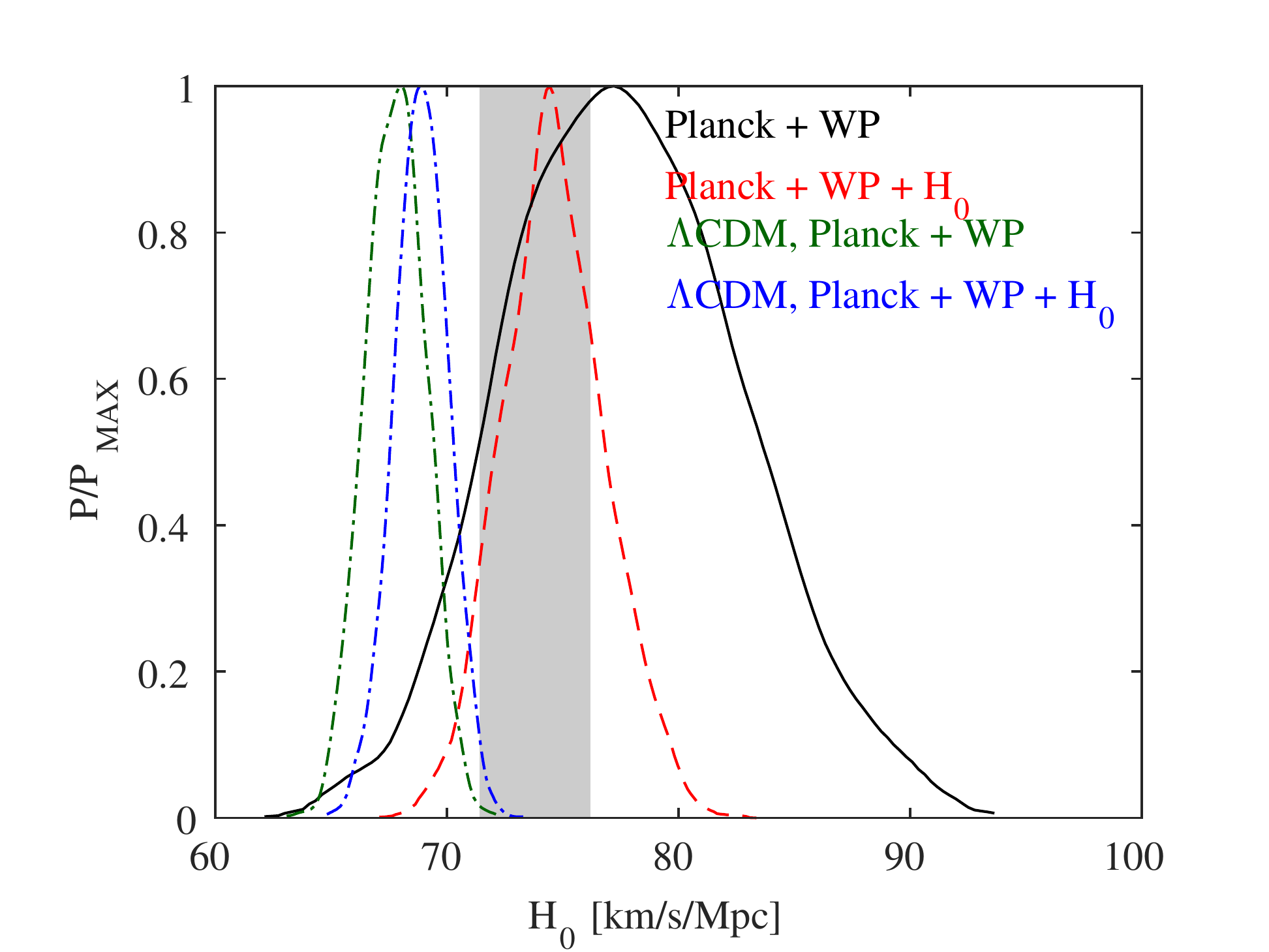}%
\vspace{0.01cm}
\caption{1D marginalised posteriors for $\Neff$ ({\it Top panel}) and $H_0$ ({\it Bottom panel}) obtained by assuming the pseudoscalar scenario and using only CMB data (black/solid line) and CMB data plus the $H_0$ prior (red/dotted line). ({\it Top panel}) The green dash-dot line refers to the $\Lambda$CDM model ($\Neff=3.046$) and the purple line is the complete thermalization case ($\Neff \simeq 4$). ({\it Bottom panel}) The green and the blue dash-dot lines show the posteriors obtained in the $\Lambda$CDM model using Planck and Planck+$H_0$, respectively. The $H_0$ prior is marked by the grey shaded region \cite{Riess:2011yx}.}%
\label{fig:contours}%
\end{figure}

In the top panel of Fig.~\ref{fig:contours} we show the 1D marginalised posterior for $\Neff$ for the Planck+WP data, as well as for the same data, but with the direct measurement of $H_0$ from \cite{Riess:2011yx} included. The data shows a clear preference for high values of $\Neff$ and the most extreme case with complete thermalisation of the sterile neutrino, corresponding to $\Neff \simeq 4$, is well within the $1\sigma$ allowed region. It is also of interest to compare the difference in $\chi^2$ between this model and the standard $\Lambda$CDM cosmology. We find that $\Delta \chi^2$ of the pseudoscalar model compared to the reference $\Lambda$CDM model is $\Delta \chi^2 = \chi^2_{\textrm{pseudoscalar}} - \chi^2_{\Lambda \textrm{CDM}} =0.298$, while if we assume $\Neff \simeq 4$  $\Delta \chi^2=0.276$.

Interestingly for this model with a subdominant, strongly interacting neutrino sector we also find a preference for a higher value of $H_0$. This effect was seen already in \cite{Hannestad:2004qu} but with a much more dramatic increase in $H_0$ because all neutrinos were assumed to be strongly interacting. In the bottom panel of Fig.~\ref{fig:contours} we show the 1D marginalised posterior for $H_0$ for this model as well as for $\Lambda$CDM. The increase in $H_0$ alleviates the tension between the locally measured value of $H_0$ and the much lower value inferred from Planck data when the standard model is assumed. We see this effect very directly when comparing $\chi^2$ values: $\Delta \chi^2 = \chi^2_{\textrm{pseudoscalar}} - \chi^2_{\Lambda \textrm{CDM}} = - 3.752$, while if we assume $\Neff \simeq 4$  $\Delta \chi^2=-3.248$. We thus find that in this case the model with a strongly interacting $\nu_s-\phi$ sector is a {\it better} fit to current data than $\Lambda$CDM (and of course a vastly better fit than $\Lambda$CDM with an additional 1 $\electronvolt$ sterile neutrino).

{\it Dark matter. ---}
We will now investigate the possibility that dark matter also couples to the new pseudo-scalar with a dimensionless coupling strength, $g_d$. We assume that the dark matter is produced at a very high temperature by e.g. inflaton decay. Once dark matter is coupled to the new interaction, there is the potential worry that it will pair annihilate via the process $\chi \bar \chi \to \phi \phi$ with the same cross section as in Eq.~(\ref{eq:pair}). If the annihilation process is in equilibrium where $\chi$ goes non-relativistic, it will dilute the density of $\chi$ while transferring an unacceptable amount of entropy to $\phi$. Due to the nature of the interaction, it is decoupled at high temperatures, and the cross section likewise drops when the dark matter becomes non-relativistic.
Therefore, we only need to ensure that the dark matter annihilation rate is low enough at $T_{\mathrm{max}} \sim m_\chi$. We assume that the cross section is given by the highly relativistic expression for $\left< \sigma |v| \right>$ in Eq.~(\ref{eq:pair}), and use the condition $\Gamma(T_{\mathrm{max}}) = \left< \sigma |v| \right> n_\chi < H(T_{\mathrm{max}})$ to derive the condition,
\begin{equation}
g_d \lesssim 2 \times 10^{-5} \left(\frac{m_\chi}{\mega\electronvolt }\right)^{1/4},
\label{eq:thermalise}
\end{equation}
for the new interaction not to overly dilute the density of $\chi$.

Additionally, the new coupling also induces a Yukawa type potential between the dark matter particles. This in turn leads to dark matter self-interactions which might have observable consequences for galactic dynamics. Rather than going through a detailed calculation we will simply estimate the mean time between dark matter scatterings in order to estimate whether self-interactions are important. In order to do so we will follow the prescription given in \cite{Ackerman:mha}.
First, following Ref.~\cite{Bellazzini:2013foa} we write
\begin{equation} \label{eq:DMpotential}
V(r) = -\frac{g_d^2}{m_\chi^2} \frac{e^{-m_\phi r}}{4\pi r^3} h(m_\phi r) \mathcal{S}, 
\end{equation}
where $h(m_\phi,r) = 1 + m_\phi r + \frac{1}{3} (m_\phi r)^2$ and $\mathcal{S}$ is a spin-dependent factor which we assume to be one. 

The interaction potential in Eq.~(\ref{eq:DMpotential}) causes elastic scattering of dark matter, and following the prescription in \cite{Ackerman:mha} we can estimate the value of $g_d$ needed in order to have a significant impact on galactic dynamics. The calculation in \cite{Ackerman:mha} was performed for a massless $U(1)$ vector so the potential is Coulomb-like. This in turn leads to both ``soft'' and ``hard'' scattering of roughly equal importance. Here we can safely neglect the contribution from soft scatterings because of the steepness of the potential. 

The ratio of the scattering time scale $\tau_\text{scat.}$ to the dynamical time scale in the galaxy $\tau_\text{dyn.}$ is given by Eq.~17 in \cite{Ackerman:mha},
\begin{equation}
\frac{\tau_\text{scat.}}{\tau_\text{dyn.}} = \frac{2R^2}{3N\sigma},
\end{equation}
where $R$ is the radius of the galaxy, $N$ is the number of DM particles in the galaxy and $\sigma$ is the scattering cross section. For a hard scatter we have $\sigma \simeq b^2$ where the impact parameter $b$ is the radial distance such that the sum of kinetic and potential energy is zero,
\begin{equation}
\frac{\alpha_d}{m_\chi^2 b^3} = \frac{1}{2} m_\chi v^2,
\end{equation}
where we have used that $m_\phi b \sim m_\phi/m_\chi \ll 1$ which leads to the approximation $V(r) \approx -\alpha_d /(m_\chi^2 r^3)$ where $\alpha_d = g_d^2/4\pi$. We then find that
\begin{equation}
\left( \frac{\tau_\text{scat.}}{\tau_\text{dyn.}} \right)^3 = \frac{2R^4 m_\chi^8 G^2}{27 N \alpha_d^2},
\end{equation}
where $G$ is Newton's constant. The condition for the time scale of scattering to be less than the age of the Universe is\footnote{We take $\tau_\text{dyn.}$ to be the dynamical time scale of a Milky Way size halo.} $\tau_\text{scat.}/\tau_\text{dyn.} \lesssim 50$. Plugging in numbers for a Milky Way size halo and using $\alpha_d = g_d^2/4\pi$, we find
\begin{equation}
g_d \gtrsim 6 \times 10^{-8} \left( \frac{m_\chi}{\mega\electronvolt} \right)^\frac{9}{4}.
\label{eq:hard}
\end{equation}
The value of $g_d$ in Eq.~(\ref{eq:hard}) can be seen as a lower bound on the value required to have a significant effect. The actual value required might be somewhat larger.

In order for elastic scattering to be important in itself the mass of the dark matter particle is therefore required to be quite small. For example, $g_d \sim 10^{-5}$ leads to the requirement that $m_\chi \lesssim 10~\mega\electronvolt$. So depending on the unknown mass of the Dark Matter particle, hard scattering on this potential could have a \emph{direct} impact on galactic dynamics. Even if this is not the case, the potential could still have a very important \emph{indirect} effect through the Sommerfeld mechanism~\cite{Bellazzini:2013foa}. The idea is that the Dark Matter particles could have some weak short range scattering cross section generated by beyond the standard model (BSM) physics, which is then enhanced by a velocity dependent boost factor $S(v)$ such that $\sigma(v) = S(v) \sigma_0$. If this new BSM physics enters at a scale $\Lambda_\text{BSM}$, we could expect $\sigma_0 \sim 1/\Lambda_\text{BSM}^2$.

{\it Sommerfeld enhanced scattering. ---}
The potential in Eq.~\eqref{eq:DMpotential} diverges faster than $r^{-2}$ so it is singular and leads to an unbounded Hamilton operator~\cite{Bedaque:2009ri}. This is of course not physical, since the potential will ultimately be regularised by UV physics. While the boost factor can be made independent of the regularisation procedure, it will depend a bit on the UV completion~\cite{Bedaque:2009ri,Bellazzini:2013foa}. We are just trying to estimate this effect, so we follow the simplified version of the regularisation procedure outlined in~\cite{Bellazzini:2013foa}: we introduce a cut-off in the potential defined by $V(r_\text{cut}) = \Lambda_\text{BSM}$ and set $V(r<r_\text{cut}) \equiv V(r_\text{cut})$ such that the potential is continuous at $r_\text{cut}$.

To compute the Sommerfeld factor, we follow~\cite{Bellazzini:2013foa} and write the radial part of the Schr\"{o}dinger equation as
\begin{align} \label{eq:schrodinger}
\Phi^{\prime \prime}_{\ell} (x) &= \left( \frac{m_\chi}{p^2} V\left( \frac{x}{p} \right) + \frac{\ell(\ell+1)}{x^2} - 1 \right)  \Phi_{\ell} (x), \\
&=  \left( \frac{-g_d^2 v}{8\pi x_m^3} h(F x_m) e^{-F x_m} + \frac{\ell(\ell+1)}{x^2} - 1 \right)  \Phi_{\ell} (x). \nonumber
\end{align}
with $x\equiv p r$ and $F\equiv \frac{2 m_\phi}{m_\chi v}$. The continuous box renormalisation has been implemented by simply using $x_m\equiv \text{max}(x, x_\text{cut})$ inside the potential term. The equation determining the cutoff $x_\text{cut}$ is
\begin{equation} \label{eq:xcut}
1 = \left( \frac{m_\chi}{\Lambda_\text{BSM}} \right) \frac{g_d^2 v^3}{32\pi x_{\textrm{cut}}^3} h(F x_{\textrm{cut}}) e^{-F x_{\textrm{cut}}}.
\end{equation}

In the limit $x \rightarrow 0$, the complete solution to Eq.~\eqref{eq:schrodinger} are $A x^{\ell + 1} + B x^{-\ell}$ for $\ell \geq 0$. As usual, requiring the solution to be regular at $x=0$ forces $B=0$. $A$ can be absorbed into the overall normalisation of the wave function, i.e. we put $A=1$. In the asymptotic limit $x\rightarrow \infty$, the solution just becomes a sine with an amplitude and a phase shift. We have
\begin{align}
\Phi_{\ell} (x) &\rightarrow x^{\ell+1}, & x\rightarrow 0, \label{eq:philimzero}\\ 
\Phi_{\ell} (x) &\rightarrow C \sin ( x - \ell \pi/2 +\delta_\ell), & x\rightarrow \infty. \label{eq:philiminfty}
\end{align}

\begin{figure}
\includegraphics[width=\columnwidth]{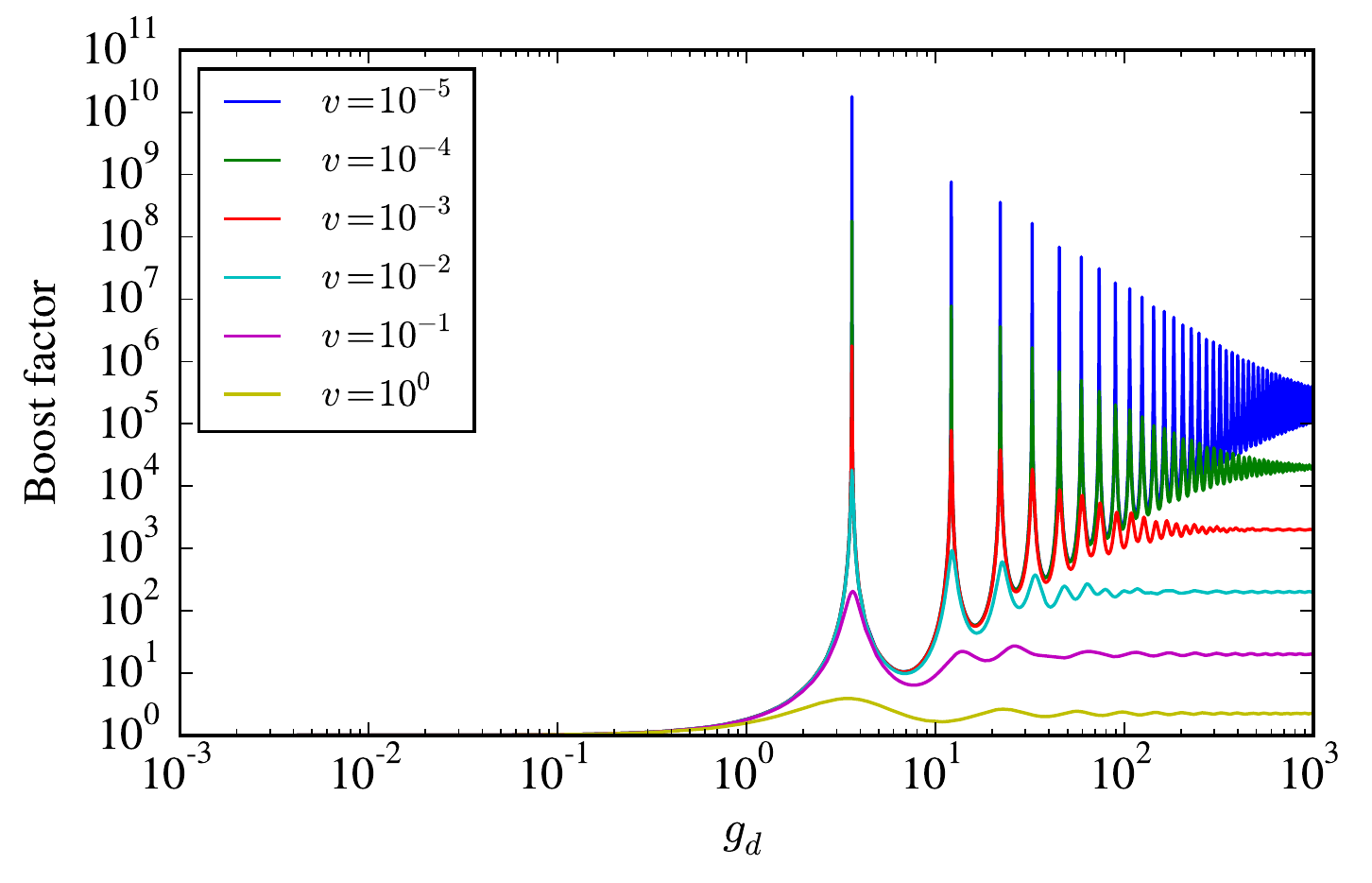}%

\includegraphics[width=\columnwidth]{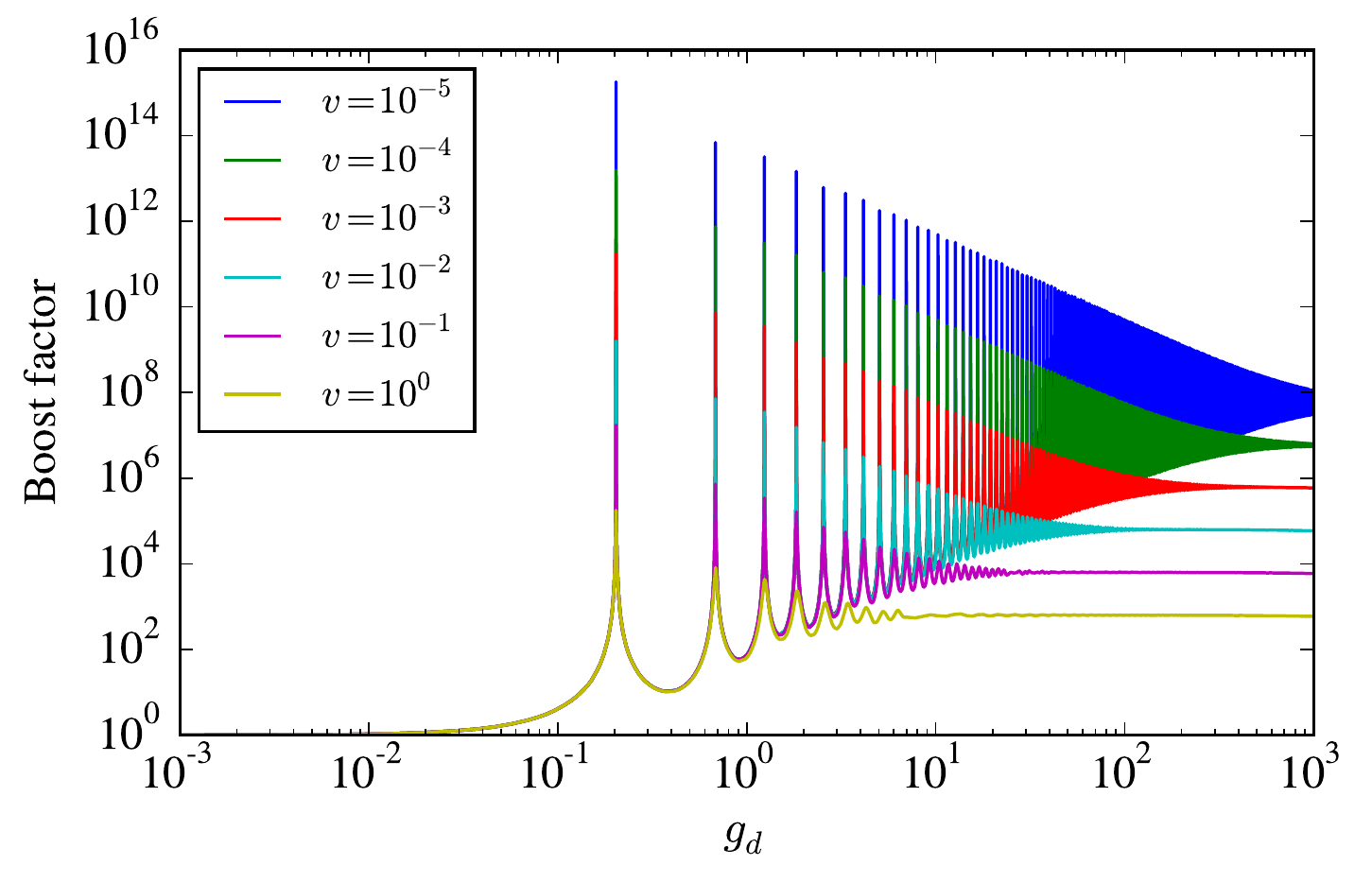}
\caption{Sommerfeld enhancement factor for $\ell=0$ due the potential in Eq.~\eqref{eq:DMpotential} for two extreme values of the ratio $(m_\chi/\Lambda_\text{BSM})$. {\it Top panel:} $(m_\chi/\Lambda_\text{BSM})=1.0$. {\it Bottom panel:} $(m_\chi/\Lambda_\text{BSM})=10^{-5}$. As discussed in the text, the dependence on the ratio $(m_\phi/m_\chi)$ is negligible.}%
\label{fig:Sommerfeld}%
\end{figure}

To compute the Sommerfeld factor numerically, we use Eq.~\eqref{eq:philimzero} to set initial conditions at $x_\text{ini}$, $0 < x_\text{ini} < x_\text{cut}$. We then evolve the wave until it has reached its asymptote in Eq.~\eqref{eq:philiminfty} and we denote this point by $x_\text{asym.}$. This happens when the wave no longer feels the potential and, for $\ell > 0$, the centrifugal barrier. The Sommerfeld factor is related to the asymptotic amplitude $C$ (through the overall normalisation) by the formula~\cite{Bellazzini:2013foa}
\begin{equation}
S_\ell = \frac{\left[(2\ell+1)!!\right]^2}{C^2} = \frac{\left[(2\ell+1)!!\right]^2}{\Phi_{\ell}^2(x_\text{asym.}) + \Phi^{\prime 2}_{\ell} (x_\text{asym.})}.
\end{equation}
The last expression is obtained from Eq.~\eqref{eq:philiminfty} and is numerically convenient. Note that the equation for the boost factor does not depend on the masses but only on $\Phi_\ell$. The mass dependence in Eq.~\eqref{eq:schrodinger} enters only through the ratio $m_\phi/m_\chi$ in the factor $h(Fx) e^{-Fx}$. This factor is $\sim 1$ when $Fx \lesssim 1$, and it is easy to show that this is the case for all values of $x$ where the potential is non-negligible, provided that $\frac{m_\phi}{m_\chi} < (v/g_d)^{\frac{2}{3}}$. This inequality is easily satisfied for the parameter space that we are considering. The regularisation procedure introduces another possible mass dependence through Eq.~\eqref{eq:xcut}. The previous argument applies again to the factor $h(Fx) e^{-Fx}$, ruling out a dependence on the $(m_\phi/m_\chi)$-ratio. So the only mass dependence will enter through the ratio $(m_\chi/\Lambda_\text{BSM})$. We have shown the boost factor in Fig.~\ref{fig:Sommerfeld} for two extreme values of this ratio. Evidently, the effect of Sommerfeld enhancement can be safely neglected for all reasonable values of $g_d$.

{\it Dark acoustic oscillations? ---}
Since our model couples dark matter to a background of dark radiation we might worry that the $\chi-\phi$ system can undergo acoustic oscillations close to the epoch of recombination and thus distort the observed CMB spectrum (see e.g.\ \cite{Cyr-Racine:2013fsa} for a recent discussion).
The interaction around the epoch of CMB formation is primarily Compton scattering, $\chi \phi \to \chi \phi$, and we can directly compare it to the normal Compton scattering rate of photons and electrons. The Compton cross section scales as $\sigma \propto \alpha^2/m^2$ where $m$ is the fermion mass. As long as $g_d^2 \ll \alpha$ and $m_\chi \gg m_e$, the dark sector acoustic oscillations will be completely negligible and therefore cosmologically safe.
This of course also means that late-time Compton scatterings can be safely ignored since they have no impact on the ability of $\chi$ to cluster gravitationally. Scaling relative to the electron-photon process we can formulate the bound as
\begin{equation}
g_d^2 \ll 1.6 \times 10^{-2} \left(\frac{m_\chi}{\mega\electronvolt}\right).
\end{equation}

{\it Discussion. ---}
We have studied a model with secret sterile neutrino interactions mediated by a massless or very light pseudoscalar. The model has some of the same features as the previously studied models based on Fermi-like interactions mediated by heavy vector bosons in the sense that it provides a background potential which can block the production of sterile neutrinos and resolve the apparent inconsistency between cosmology and short baseline neutrino oscillation data.

However, the model has very different late-time phenomenology. The very low mass of the pseudoscalar makes the sterile neutrino strongly self-interacting at late times, an effect which is perfectly consistent with current cosmological data, but might be used to uniquely identify the model once more precise measurements become available.
In order to accommodate the mass bound from cosmological large scale structure~\cite{Mirizzi:2014ama}, we need $g_s \gtrsim 10^{-6}$ to allow the sterile neutrinos to annihilate when they become non-relativistic. Our analysis of the CMB suggests $\Neff \approx 4$, and this suggestion is amplified if we also consider the direct measurements of $H_0$. At $95\%$ confidence we can rule out $\Neff = 3.046$ when we include the H0 measurement, and this formally corresponds to an upper limit on $g_s$ of $g_s \lesssim 10^{-5}$ according to Fig.~\ref{fig:Neff}. However, this bound is very dependent on the set of data we have used, and might both be strengthened and weakened by including more data. 
We finally arrive at a combined bound on $g_s$ of
\begin{equation}
  \label{eq:gsbounds}
  10^{-6} \lesssim g_s \lesssim 10^{-5} \text{(CMB+H0). } 
\end{equation}
A more robust determination of $\Neff$ would allow the possible values for $g_s$ to be further confined, and a precise value of $\Neff > 3.046$ would allow us to pinpoint a corresponding coupling strength.
We also note that since the fundamental coupling strength is very low and restricted to the sterile sector in this model it is unlikely to produce observable effects on neutrino physics in general (see e.g.\ \cite{Bernatowicz:1992ma} for laboratory constraints).
Considering non-standard energy loss from the proto-neutron star in SN1987a also leads to an upper bound on $g_s$ in the $\sim {\rm few} \times 10^{-5}$ range (see e.g.\ \cite{Raffelt:1996wa} for a discussion).

In addition to the coupling to sterile neutrinos we hypothesise that the pseudoscalar also couples to the dark matter particle. Provided that the dark matter particle is sufficiently light this can lead to significant effects on dark matter clustering in galaxies and clusters and possibly resolve some of the apparent discrepancies between the standard $\Lambda$CDM model and observations~\cite{Vogelsberger:2012ku}. These discrepancies include the ``Too big to fail'' problem~\cite{BoylanKolchin:2011de} and the ``cusp vs. core'' problem (see~\cite{deBlok:2009sp} and references herein), but not the ``missing satellites'' problem~\cite{Klypin:1999uc} which would require a stronger coupling between neutrinos and DM.

\begin{figure}
\includegraphics[width=\columnwidth]{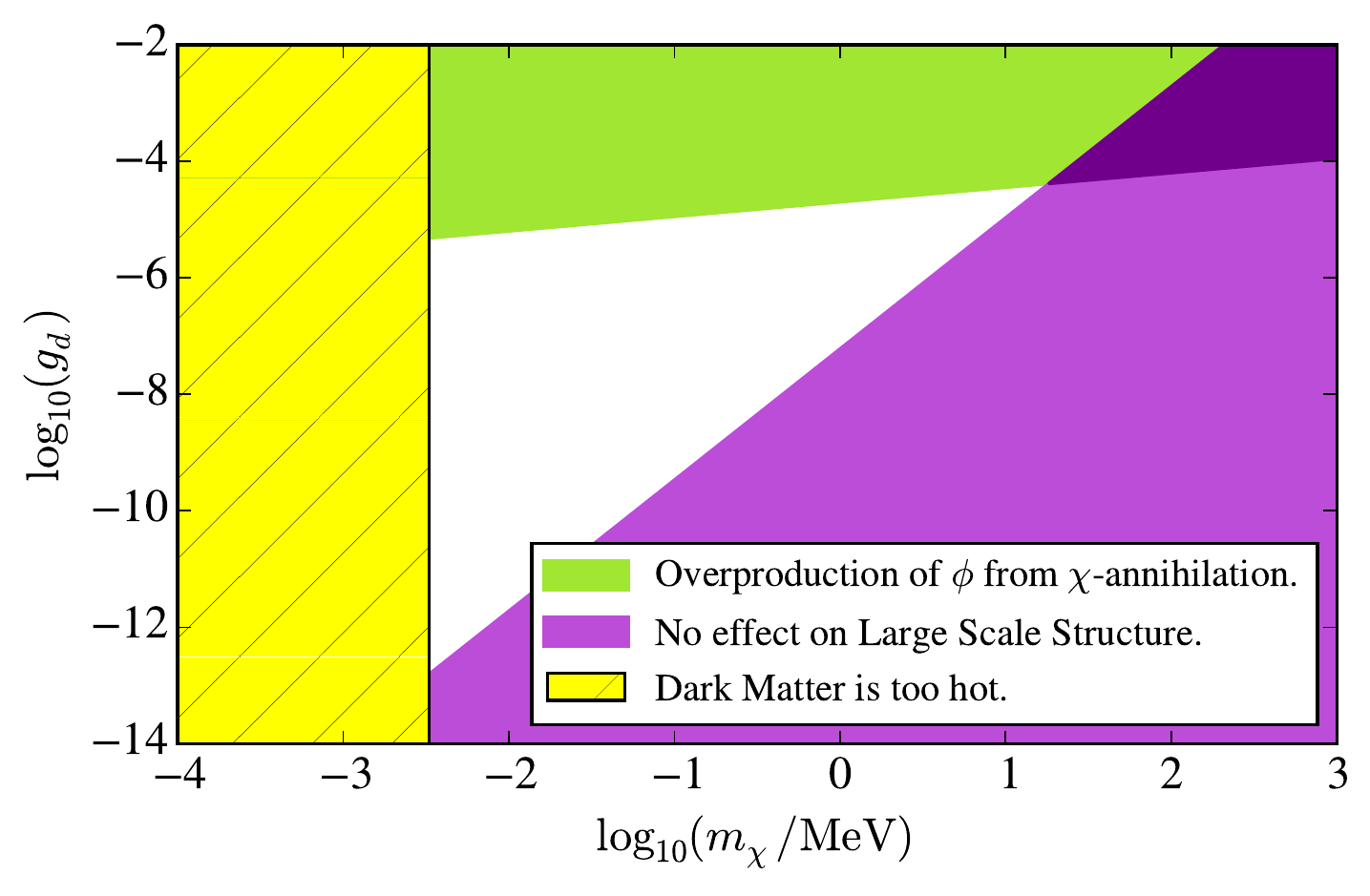}%
\caption{Constraints in $m_\chi-g_d$ space. The green region is ruled out from Eq.~(\ref{eq:thermalise}) due to overproduction of $\phi$-particles from $\chi$-annihilations, while the purple region will have no effect on galactic dynamics, cf. Eq.~(\ref{eq:hard}).}%
\label{fig:constraints}
\end{figure}

In order for the model to be viable, the dark matter coupling must be sufficiently low that the pair annihilations do not transfer excess entropy to the plasma of sterile neutrinos and pseudoscalars. Conversely, the dark matter coupling must be strong enough to produce an observable effect on galactic dynamics. In Fig.~\ref{fig:constraints} we show these two constraints simultaneously and include the bound from warm dark matter~\cite{Viel:2013fqw}. We are left with a viable DM candidate with a mass between ${\rm few}~\kilo\electronvolt$ and $\sim 10~\mega\electronvolt$ and couplings from $10^{-13}$ to $10^{-5}$. A more detailed treatment of the ``cusp vs. core'' and ``Too big to fail'' problems could probably constrain the dark matter further, but that is beyond the scope of this article. The type of dark matter, that we have described, is very different from the normal WIMP cold dark matter. However, it is entirely possible that dark matter consists of an additional sterile neutrino species with extremely suppressed mixing to the active sector. If this is the case it cannot be produced via the usual scattering and oscillation mechanism. However, unlike an $\mega\electronvolt$ sterile neutrino produced via the normal oscillation and scattering mechanism it also remains stable on cosmological timescales.
The actual production mechanism for the dark matter particle might be via direct inflaton decay at reheating or from the thermal background at very high temperature. 

In summary, sterile neutrino and dark matter interactions via a light pseudoscalar seems an extremely interesting possibility for explaining a variety of different problems in cosmology and certainly merits further study.

{\it Acknowledgments.}---
MA acknowledges partial support from the European Union FP7 ITN INVISIBLES (Marie Curie Actions, PITN- GA-2011- 289442)

%%%%%%%%%%%%%%%%%%%%%%%%%%%%%%%%%%%%%%%%%%%%%%%%%%%%%%%%%%%%%%%%%%%%%%%

%%%%%%%%%%%%%%%%%%%%%%%%%%%%%%%%%%%%%%%%%%%%%%%%%%%%%%%%%%%%%%%%%%%%%%%

\begin{thebibliography}{99}
%%%%%%%%%%%%%%%%%%%%%%%%%%%%%%%%%%%%%%%%%%%%%%%%%%%%%%%%%%%%%%%%%%%%%%%


%
\bibitem{Kopp:2013vaa} 
  %J.~Kopp, P.~A.~N.~Machado, M.~Maltoni and T.~Schwetz,
  J.~Kopp {\it et al.},
  %``Sterile Neutrino Oscillations: The Global Picture,''
  JHEP {\bf 1305}, 050 (2013)
  %[arXiv:1303.3011 [hep-ph]].
  %%CITATION = ARXIV:1303.3011;%%
  %38 citations counted in INSPIRE as of 11 Oct 2013
	
%\cite{Kopp:2013vaa,Giunti:2012bc}
\bibitem{Giunti:2012bc} 
  C.~Giunti, M.~Laveder, Y.~F.~Li and H.~W.~Long,
  %C.~Giunti {\it et al.},
	%``Short-Baseline Electron Neutrino Oscillation Length After Troitsk,''
  Phys.\ Rev.\ D {\bf 87}, 013004 (2013)
  %[arXiv:1212.3805 [hep-ph]].
  %%CITATION = ARXIV:1212.3805;%%
  %7 citations counted in INSPIRE as of 11 Oct 2013

\bibitem{Enqvist:1991qj} 
  K.~Enqvist, K.~Kainulainen and M.~J.~Thomson,
  %``Stringent cosmological bounds on inert neutrino mixing,''
  Nucl.\ Phys.\ B {\bf 373}, 498 (1992).
  %%CITATION = NUPHA,B373,498;%%
  %240 citations counted in INSPIRE as of 10 Apr 2014
	
%\cite{Hannestad:2012ky}
\bibitem{Hannestad:2012ky} 
  S.~Hannestad, I.~Tamborra and T.~Tram,
  %``Thermalisation of light sterile neutrinos in the early universe,''
  JCAP {\bf 1207}, 025 (2012).
  %[arXiv:1204.5861 [astro-ph.CO]].
  %%CITATION = ARXIV:1204.5861;%%
  %30 citations counted in INSPIRE as of 02 Oct 2013
	
	%\cite{Saviano:2013ktj}
\bibitem{Saviano:2013ktj} 
  N.~Saviano {\it et al.},
	%N.~Saviano, A.~Mirizzi, O.~Pisanti, P.~D.~Serpico, G.~Mangano and G.~Miele,
  %``Multi-momentum and multi-flavour active-sterile neutrino oscillations in the early universe: role of neutrino asymmetries and effects on nucleosynthesis,''
  Phys.\ Rev.\ D {\bf 87}, 073006 (2013).
  %%CITATION = ARXIV:1302.1200;%%
  %6 citations counted in INSPIRE as of 10 Oct 2013
		

%\cite{Archidiacono:2014apa}
\bibitem{Archidiacono:2014apa} 
  M.~Archidiacono, N.~Fornengo, S.~Gariazzo, C.~Giunti, S.~Hannestad and M.~Laveder,
  %``Light sterile neutrinos after BICEP-2,''
  arXiv:1404.1794 [astro-ph.CO].
  %%CITATION = ARXIV:1404.1794;%%
	
%\cite{Ade:2013zuv}
\bibitem{Ade:2013zuv} 
  P.~A.~R.~Ade {\it et al.},
  %``Planck 2013 results. XVI. Cosmological parameters,''
  arXiv:1303.5076
  %%CITATION = ARXIV:1303.5076;%%
  %696 citations counted in INSPIRE as of 02 Oct 2013
	
\bibitem{Dvorkin:2014lea} 
  C.~Dvorkin, M.~Wyman, D.~H.~Rudd and W.~Hu,
  %``Neutrinos help reconcile Planck measurements with both Early and Local Universe,''
  arXiv:1403.8049 [astro-ph.CO].
  %%CITATION = ARXIV:1403.8049;%%
  %5 citations counted in INSPIRE as of 10 Apr 2014
	
	%\cite{Zhang:2014dxk}
\bibitem{Zhang:2014dxk} 
  J.~-F.~Zhang, Y.~-H.~Li and X.~Zhang,
  %``Sterile neutrinos help reconcile the observational results of primordial gravitational waves from Planck and BICEP2,''
  arXiv:1403.7028 [astro-ph.CO].
  %%CITATION = ARXIV:1403.7028;%%
  %9 citations counted in INSPIRE as of 10 Apr 2014
	%
	
	
	\bibitem{Hannestad:2013ana}
  S.~Hannestad, R.~S.~Hansen and T.~Tram,
  %``How secret interactions can reconcile sterile neutrinos with cosmology,''
  Phys.\ Rev.\ Lett.\  {\bf 112} (2014) 031802
  [arXiv:1310.5926 [astro-ph.CO]].
  %%CITATION = ARXIV:1310.5926;%%
  %5 citations counted in INSPIRE as of 02 Apr 2014
	
%
\bibitem{Dasgupta:2013zpn}
  B.~Dasgupta and J.~Kopp,
  %``A m\'enage \`a trois of eV-scale sterile neutrinos, cosmology, and structure formation,''
  Phys.\ Rev.\ Lett.\  {\bf 112} (2014) 031803
  [arXiv:1310.6337 [hep-ph]].
  %%CITATION = ARXIV:1310.6337;%%
  %7 citations counted in INSPIRE as of 02 Apr 2014
	
%\cite{Bringmann:2013vra}
\bibitem{Bringmann:2013vra}
  T.~Bringmann, J.~Hasenkamp and J.~Kersten,
  %``Tight bonds between sterile neutrinos and dark matter,''
  arXiv:1312.4947 [hep-ph].
  %%CITATION = ARXIV:1312.4947;%%
  %1 citations counted in INSPIRE as of 02 Apr 2014

%\cite{Ko:2014bka}
\bibitem{Ko:2014bka} 
  P.~Ko and Y.~Tang,
  %``$\nu \Lambda$MDM: A Model for Sterile Neutrino and Dark Matter Reconciles Cosmological and Neutrino Oscillation Data after BICEP2,''
  arXiv:1404.0236 [hep-ph].
  %%CITATION = ARXIV:1404.0236;%%
  %4 citations counted in INSPIRE as of 30 Apr 2014
	%\cite{Rehagen:2014vna}

\bibitem{Rehagen:2014vna}
  T.~Rehagen and G.~B.~Gelmini,
  %``Effects of kination and scalar-tensor cosmologies on sterile neutrinos,''
  arXiv:1402.0607 [hep-ph].
  %%CITATION = ARXIV:1402.0607;%%
	%

\bibitem{Archidiacono:2013dua} 
  M.~Archidiacono and S.~Hannestad,
  %``Updated constraints on non-standard neutrino interactions from Planck,''
  arXiv:1311.3873 [astro-ph.CO].
  %%CITATION = ARXIV:1311.3873;%%

%\cite{Farzan:2002wx}
\bibitem{Farzan:2002wx} 
  Y.~Farzan,
  %``Bounds on the coupling of the Majoron to light neutrinos from supernova cooling,''
  Phys.\ Rev.\ D {\bf 67}, 073015 (2003)
  [hep-ph/0211375].
  %%CITATION = HEP-PH/0211375;%%
  %43 citations counted in INSPIRE as of 15 Jan 2015

%\cite{Kachelriess:2000qc}
\bibitem{Kachelriess:2000qc} 
  M.~Kachelriess, R.~Tomas and J.~W.~F.~Valle,
  %``Supernova bounds on Majoron emitting decays of light neutrinos,''
  Phys.\ Rev.\ D {\bf 62}, 023004 (2000)
  [hep-ph/0001039].
  %%CITATION = HEP-PH/0001039;%%

%\cite{Bernatowicz:1992ma}
\bibitem{Bernatowicz:1992ma} 
  T.~Bernatowicz, J.~Brannon, R.~Brazzle, R.~Cowsik, C.~Hohenberg and F.~Podosek,
  %``Neutrino mass limits for a precise determination of beta-beta decay rates of Te-128 and Te-130,''
  Phys.\ Rev.\ Lett.\  {\bf 69}, 2341 (1992).
  %%CITATION = PRLTA,69,2341;%%

	%\cite{Babu:1991at}
\bibitem{Babu:1991at} 
  K.~S.~Babu and I.~Z.~Rothstein,
  %``Relaxing nucleosynthesis bounds on sterile-neutrinos,''
  Phys.\ Lett.\ B {\bf 275}, 112 (1992).
  %%CITATION = PHLTA,B275,112;%%
  %24 citations counted in INSPIRE as of 03 Dec 2013

%\cite{Enqvist:1992}
\bibitem{Enqvist:1992} 
  K.~Enqvist, K.~Kainulainen and M.~J.~Thomson,
  %``Cosmological bounds on Dirac-Majorana neutrinos,''
  Phys.\ Lett.\ B {\bf 280}, 245 (1992).
  %%CITATION = PHLTA,B280,245;%%
  %19 citations counted in INSPIRE as of 03 Dec 2013




%
	
%\cite{Enqvist:1990ad}
\bibitem{Enqvist:1990ad} 
  K.~Enqvist, K.~Kainulainen and J.~Maalampi,
  %``Refraction and Oscillations of Neutrinos in the Early Universe,''
  Nucl.\ Phys.\ B {\bf 349}, 754 (1991).
  %%CITATION = NUPHA,B349,754;%%
  %145 citations counted in INSPIRE as of 21 Oct 2013

\bibitem{McKellar:1992ja}
B.H.J.~McKellar and M.J.~Thomson,
PRD {\bf 49}, 2710 (1994).


\bibitem{Sigl:1992fn}
G.~Sigl and G.~Raffelt,
Nucl. Phys. B {\bf 406}, 423 (1993).
			


	%
\bibitem{Stodolsky:1986dx} 
  L.~Stodolsky,
  %``On the Treatment of Neutrino Oscillations in a Thermal Environment,''
  Phys.\ Rev.\ D {\bf 36}, 2273 (1987).
  %%CITATION = PHRVA,D36,2273;%%
  %143 citations counted in INSPIRE as of 10 Oct 2013
	
	


%\cite{Dolgov:1996fp}
\bibitem{Dolgov:1996fp}
  A.~D.~Dolgov, S.~Pastor, J.~C.~Romao and J.~W.~F.~Valle,
  %``Primordial nucleosynthesis, Majorons and heavy tau neutrinos,''
  Nucl.\ Phys.\ B {\bf 496} (1997) 24
  [hep-ph/9610507].
  %%CITATION = HEP-PH/9610507;%%
  %28 citations counted in INSPIRE as of 22 Jan 2014

%\cite{Mirizzi:2012we}
\bibitem{Mirizzi:2012we}
  A.~Mirizzi, N.~Saviano, G.~Miele and P.~D.~Serpico,
  %``Light sterile neutrino production in the early universe with dynamical neutrino asymmetries,''
  Phys.\ Rev.\ D {\bf 86} (2012) 053009
  [arXiv:1206.1046 [hep-ph]].
  %%CITATION = ARXIV:1206.1046;%%
  %24 citations counted in INSPIRE as of 15 Aug 2014

\bibitem{Hannestad:2013pha}
  S.~Hannestad, R.~S.~Hansen and T.~Tram,
%      title          = "{Can active-sterile neutrino oscillations lead to chaotic
%                        behavior of the cosmological lepton asymmetry?}",
  JCAP {\bf 1304}, 032 (2013).
%      journal        = "JCAP",
%      volume         = "1304",
%      pages          = "032",
%      doi            = "10.1088/1475-7516/2013/04/032",
%      year           = "2013",
%      eprint         = "1302.7279",
%      archivePrefix  = "arXiv",
%      primaryClass   = "astro-ph.CO",
%      SLACcitation   = "
%%CITATION = ARXIV:1302.7279;%%",
%}

\bibitem{Kainulainen:2001cb}
K.~Kainulainen and A.~Sorri,
JHEP {\bf 0202}, 020 (2002).
%      volume         = "0202",
%      pages          = "020",
%      year           = "2002",
%      eprint         = "hep-ph/0112158",
%      archivePrefix  = "arXiv",
%      primaryClass   = "hep-ph",
%      reportNumber   = "NORDITA-2001-70-HE",
%      SLACcitation   = "%%CITATION = HEP-PH/0112158;%%",


%
\bibitem{Barbieri:1990vx} 
  R.~Barbieri and A.~Dolgov,
  %``Neutrino oscillations in the early universe,''
  Nucl.\ Phys.\ B {\bf 349}, 743 (1991).
  %%CITATION = NUPHA,B349,743;%%
  %195 citations counted in INSPIRE as of 21 Oct 2013
	
%%%%%%%%% interactions %%%%%%%%%%%%%%%%%%%%%%%%%

%\cite{Mirizzi:2014ama}
\bibitem{Mirizzi:2014ama} 
  A.~Mirizzi, G.~Mangano, O.~Pisanti and N.~Saviano,
  %``Collisional production of sterile neutrinos via secret interactions and cosmological implications,''
  arXiv:1410.1385 [hep-ph].
  %%CITATION = ARXIV:1410.1385;%%
  %1 citations counted in INSPIRE as of 17 Dec 2014	

%\cite{Raffelt:1987ah}
\bibitem{Raffelt:1987ah} 
  G.~Raffelt and J.~Silk,
  %``Light Neutrinos As Cold Dark Matter,''
  Phys.\ Lett.\ B {\bf 192}, 65 (1987).
  %%CITATION = PHLTA,B192,65;%%
  %26 citations counted in INSPIRE as of 22 Apr 2014
	
%\cite{AtrioBarandela:1996ur}
\bibitem{AtrioBarandela:1996ur} 
  F.~Atrio-Barandela and S.~Davidson,
  %``Interacting hot dark matter,''
  Phys.\ Rev.\ D {\bf 55}, 5886 (1997)
  [astro-ph/9702236].
  %%CITATION = ASTRO-PH/9702236;%%
  %16 citations counted in INSPIRE as of 13 Nov 2013
	
	%\cite{Beacom:2004yd}
\bibitem{Beacom:2004yd} 
  J.~F.~Beacom, N.~F.~Bell and S.~Dodelson,
  %``Neutrinoless universe,''
  Phys.\ Rev.\ Lett.\  {\bf 93} (2004) 121302
  [astro-ph/0404585].
  %%CITATION = ASTRO-PH/0404585;%%
  %84 citations counted in INSPIRE as of 27 Mar 2014

%\cite{Hannestad:2004qu}
\bibitem{Hannestad:2004qu}
  S.~Hannestad,
  %``Structure formation with strongly interacting neutrinos - Implications for the cosmological neutrino mass bound,''
  JCAP {\bf 0502} (2005) 011
  [astro-ph/0411475].
  %%CITATION = ASTRO-PH/0411475;%%
  %45 citations counted in INSPIRE as of 30 Sep 2013	

%\cite{Cirelli:2006kt}
\bibitem{Cirelli:2006kt} 
  M.~Cirelli and A.~Strumia,
  %``Cosmology of neutrinos and extra light particles after WMAP3,''
  JCAP {\bf 0612}, 013 (2006)
  [astro-ph/0607086].
  %%CITATION = ASTRO-PH/0607086;%%
  %72 citations counted in INSPIRE as of 14 Nov 2013
	
	%\cite{Friedland:2007vv}
\bibitem{Friedland:2007vv} 
  A.~Friedland, K.~M.~Zurek and S.~Bashinsky,
  %``Constraining Models of Neutrino Mass and Neutrino Interactions with the Planck Satellite,''
  arXiv:0704.3271 [astro-ph].
  %%CITATION = ARXIV:0704.3271;%%
  %17 citations counted in INSPIRE as of 13 Nov 2013
	
		%\cite{Basboll:2008fx}
\bibitem{Basboll:2008fx}
  A.~Basboll, O.~E.~Bjaelde, S.~Hannestad and G.~G.~Raffelt,
  %``Are cosmological neutrinos free-streaming?,''
  Phys.\ Rev.\ D {\bf 79} (2009) 043512
  [arXiv:0806.1735 [astro-ph]].
  %%CITATION = ARXIV:0806.1735;%%
  %14 citations counted in INSPIRE as of 30 Sep 2013
	
\bibitem{Cyr-Racine:2013jua}
  F.~-Y.~Cyr-Racine and K.~Sigurdson,
  %``Limits on Neutrino-Neutrino Scattering in the Early Universe,''
  arXiv:1306.1536.
	
		%
\bibitem{Lewis:2002ah}
  A.~Lewis and S.~Bridle,
  %``Cosmological parameters from CMB and other data: a Monte-Carlo approach,''
  Phys.\ Rev.\  D {\bf 66}, 103511 (2002)
  [arXiv:astro-ph/0205436].
  %%CITATION = PHRVA,D66,103511;%%

%\cite{Ade:2013kta}
\bibitem{Ade:2013kta} 
  P.~A.~R.~Ade {\it et al.}  [Planck Collaboration],
  %``Planck 2013 results. XV. CMB power spectra and likelihood,''
  arXiv:1303.5075 [astro-ph.CO].
  %%CITATION = ARXIV:1303.5075;%%
  %248 citations counted in INSPIRE as of 13 Aug 2014


%\cite{Riess:2011yx}
\bibitem{Riess:2011yx} 
  A.~G.~Riess, L.~Macri, S.~Casertano, H.~Lampeitl, H.~C.~Ferguson, A.~V.~Filippenko, S.~W.~Jha and W.~Li {\it et al.},
  %``A 3% Solution: Determination of the Hubble Constant with the Hubble Space Telescope and Wide Field Camera 3,''
  Astrophys.\ J.\  {\bf 730}, 119 (2011)
  [Erratum-ibid.\  {\bf 732}, 129 (2011)]
  [arXiv:1103.2976 [astro-ph.CO]].
  %%CITATION = ARXIV:1103.2976;%%
  %606 citations counted in INSPIRE as of 16 Dec 2014


\bibitem{Ackerman:mha}
L.~Ackerman, M.~R.~Buckley, S.~M.~Carroll and M.~Kamionkowski,
%      title          = "{Dark Matter and Dark Radiation}",
Phys.Rev. {\bf D79} (2009) 023519
arXiv:0810.5126
%      archivePrefix  = "arXiv",
%      primaryClass   = "hep-ph",
%      SLACcitation   = "%%CITATION = ARXIV:0810.5126;%%",
%}

\bibitem{Bellazzini:2013foa}
B.~Bellazzini, M.~Cliche and P.~ Tanedo,
%      title          = "{The Effective Theory of Self-Interacting Dark Matter}",
Phys.Rev. {\bf D88} (2013) 083506
arXiv:1307.1129.
%      doi            = "10.1103/PhysRevD.88.083506",
%      year           = "2013",
%      eprint         = "1307.1129",
%      archivePrefix  = "arXiv",
%      SLACcitation   = "%%CITATION = ARXIV:1307.1129;%%",
%}	

\bibitem{Bedaque:2009ri}
P.~F.~Bedaque, M.~I.~Buchoff R.~K.~Mishra
%      title          = "{Sommerfeld enhancement from Goldstone pseudo-scalar
%                        exchange}",
JHEP {\bf 0911} (2009) 046
arXiv:0907.0235
%      archivePrefix  = "arXiv",
%      primaryClass   = "hep-ph",
%      SLACcitation   = "%%CITATION = ARXIV:0907.0235;%%",
%}

%\cite{Cyr-Racine:2013fsa}
\bibitem{Cyr-Racine:2013fsa} 
  F.~-Y.~Cyr-Racine, R.~de Putter, A.~Raccanelli and K.~Sigurdson,
  %``Constraints on Large-Scale Dark Acoustic Oscillations from Cosmology,''
  Phys.\ Rev.\ D {\bf 89}, 063517 (2014)
  [arXiv:1310.3278 [astro-ph.CO]].
  %%CITATION = ARXIV:1310.3278;%%
  %14 citations counted in INSPIRE as of 10 Apr 2014
	
%\cite{Raffelt:1996wa}
\bibitem{Raffelt:1996wa} 
  G.~G.~Raffelt,
  %``Stars as laboratories for fundamental physics : The astrophysics of neutrinos, axions, and other weakly interacting particles,''
  Chicago, USA: Univ. Pr. (1996) 664 p
  %17 citations counted in INSPIRE as of 19 Jan 2015
	
\bibitem{Vogelsberger:2012ku}
M.~Vogelsberger J.~Zavala and A.~Loeb,
%Subhaloes in Self-Interacting Galactic Dark Matter Haloes}",
Mon.Not.Roy.Astron.Soc. {\bf 423} (2012) 3740
%arXiv:1201.5892 [astro-ph.CO]

\bibitem{BoylanKolchin:2011de}
M.~Boylan-Kolchin, J.~S.~Bullock and M.~Kaplinghat,
%Too big to fail? The puzzling darkness of massive Milky Way subhaloes",
Mon.Not.Roy.Astron.Soc. {\bf 415} (2011) L40
%arXiv:1103.0007 [astro-ph.CO]

\bibitem{deBlok:2009sp}
W.~J.~G.~de Blok,
%      title          = "{The Core-Cusp Problem}",
Adv.Astron. {\bf 2010} 789293 
%arXiv:0910.3538 [astro-ph.CO]

\bibitem{Klypin:1999uc}
A.~A.~Klypin, A.~V.~Kravtsov, O.~Valenzuela and F.~Prada,
%      title          = "{Where are the missing Galactic satellites?}",
Astrophys.J. {\bf 522} (1999) 82
%astro-ph/9901240

%\cite{Viel:2013fqw}
\bibitem{Viel:2013fqw} 
  M.~Viel, G.~D.~Becker, J.~S.~Bolton and M.~G.~Haehnelt,
  %``Warm dark matter as a solution to the small scale crisis: New constraints from high redshift Lyman-α forest data,''
  Phys.\ Rev.\ D {\bf 88}, 043502 (2013)
  [arXiv:1306.2314 [astro-ph.CO]].
  %%CITATION = ARXIV:1306.2314;%%


\end{thebibliography}
\end{document}